# Dynamics of Air Transport Networks:
# A Review from a Complex Systems Perspective

Luis E C ROCHA[1,2]

[1]Department of Mathematics, Université de Namur
Rue Rempart de la Vierge 8 B5000 Namur, Belgium

[2]Department of Public Health Sciences, Karolinska Institutet
Tomtebodavägen 18A S17177 Stockholm, Sweden

luis.rocha@ki.se

May 17, 2016

## Abstract

Air transport systems are highly dynamic at temporal scales from minutes to years. This dynamic behavior not only characterizes the evolution of the system but also affect the system's functioning. Understanding the evolutionary mechanisms is thus fundamental in order to better design optimal air transport networks that benefits companies, passengers and the environment. In this review, we briefly present and discuss the state-of-art on time-evolving air transport networks. We distinguish the structural analysis of sequences of network snapshots, ideal for long-term network evolution (e.g. annual evolution), and temporal paths, preferred for short-term dynamics (e.g. hourly evolution). We emphasize that most previous research focused on the first modeling approach (i.e. long-term) whereas only a few studies look at high-resolution temporal paths. We conclude the review highlighting that much research remains to be done, both to apply already available methods and to develop new measures for temporal paths on air transport networks. In particular, we identify that the study of delays, network resilience and optimization of resources (aircraft and crew) are critical topics that can benefit of temporal network analysis.

## 1. Introduction

Air transport has been increasingly important as means of transportation in both developed and undeveloped countries [1,2]. Although associated to relatively high costs, air transport is generally safer and faster in comparison to other means of transportation [3,4] particularly to connect isolated rural areas and islands with urbanized areas, or to connect mutually distant locations such as cities in different continents. Unfortunately, air transport also contributes for the efficient spread of infectious diseases over large spatial regions [5,6]. Similarly to other modes of transport, airplanes follow pre-defined airways according to regulations of the aerial space of a given country. The collection of source-destination of flights however fundamentally characterizes the air transport network irrespective of the routes taken by aircrafts.

Altogether, pairs of cities (or airports) form a complex network of flights in which nodes represent the locations and links represent the fact that at least one flight occurred between the two locations during some interval of time [7-9]. This network perspective helps to understand the inter-connections and inter-dependencies between the multiple parts of the air transport system. On the other hand, the network framework also decreases the own complexity of air transport by reducing the model to pair-wise interactions without taking directly into account particularities of the system, as for example, impact of the weather, official regulations or types of aircrafts. This information however can be added in a sophisticated dynamic network model. Such simplifying approach is not exclusive of air transport networks but has been used in various disciplines to study the most diverse natural and man-made systems [8,9]. It helps to identify the most relevant mechanisms driving the evolution and functioning of the system. The goal is to reduce the complexity of the problem by focusing on the structure of connections between its parts. There are a number of studies focusing on the structural properties of air transport network [7,8,10]. Such studies have been increasingly appreciated by scholars studying air transport using standard methods [11]. By using network science, one is able to identify the centrality or importance of certain airports at a global or regional scale. In other words, one is able to identify bottlenecks or clusters of airports with global relevance beyond the trivial measures of accumulated traffic or size of an airport. Due to the architecture of air transport system, sometimes, medium-size airports are more strategic to connect different parts of the network than larger hub-airports [12]. These central airports may not only indicate fragile parts of the network [12], i.e. failure or attack of these airports may severely disrupt a large portion of the network, but also indicate strategic airports to implement screening and infection control in order to avoid worldwide pandemics.

One important feature of air transport networks is their dynamic structure. The timings of departure and arrival of flights vary considerable within a day, during the week or at different seasons. Given these intrinsically dynamic characteristics, the static network framework limits the study of certain properties of these networks. Although valuable insights have been provided by analyzing the static structure of flight networks at different temporal scales, there is increasing need to use more advanced methods of network science to characterize the network temporal evolution at small scales, i.e. at high temporal resolution. In this review, we introduce basic concepts of network science, particularly emphasizing temporal networks, to those not familiar with the topic. We also review available literature dealing with empirical analysis of evolving air transport networks. We will not review papers focusing on theoretical modeling of the evolution of air transport systems even if these models aim to reproduce empirical observations (e.g. refs. [13-17]). Our expectation is that this review paper fills in knowledge gaps and encourages further collaboration between traditional research and network science for better understanding of the capacities and limitations of air transport systems. Network science can be used as a quantitative supporting tool to better understand the complexity of the various layers of the air transport system.

The review is organized by first introducing fundamental concepts of network science, in particular, the definition of static and temporal networks, and basic measures used to extract information of air transport networks (sections 2.1 and 2.2). We then briefly summarize the sources of data used in the reviewed papers (section 2.3) and some computational tools for network analysis (section 2.4). These are followed by a literature review of airport networks at local (section 3) and at country (section 4) levels, and the analysis of air route networks (section 5). Afterwards, we present a few results on temporal networks (section 6) and complete the review summarizing the past and discussing perspectives for future use of temporal network methods on air transport networks (section 7).

## 2. Network Science

In this section, we first present some fundamental concepts and measures of network science applied to air transport networks. Afterwards, we define and discuss some aspects of temporal networks. The section is closed with a presentation of data sources and computational tools for network analysis.

### 2.1. Static Networks

The air transport network can be defined by a set of locations (airports, cities, regions or countries) named nodes (or vertices), and a set of links (or edges), representing the flights, connecting these locations pair-wise (Figure 1). The level of spatial aggregation, from airports to countries, depends on the research questions and interests of the investigator. Although the choice of this spatial resolution may affect the network measures, the same methods, as described bellow, may be used in any network model. Networks (mathematically described by graphs) are typically represented by an adjacency matrix $A$ of size $NxN$ (where $N$ is the number of nodes) which elements $a_{ij}$ are equal to 1 if there is at least one flight between nodes $i$ and $j$ and equal to 0 otherwise [8,9]. One may associate values, called weights, to links in order to represent features of the flights (e.g. number of passengers or duration of the flight) or features of the respective route between the locations (e.g. number of flights per day or distance). In this case, a matrix $W$, with elements $w_{ij}$, is used to describe the network. In a multiplex framework [18-20], nodes are typically fixed on each layer (though this is not necessary) and the respective links correspond to different features of the respective flight or route (e.g. number of passengers or amount of cargo transported). Note that links can be either directed ($a_{ij} \neq a_{ji}$) or undirected ($a_{ij} = a_{ji}$) according to source-destination routes.

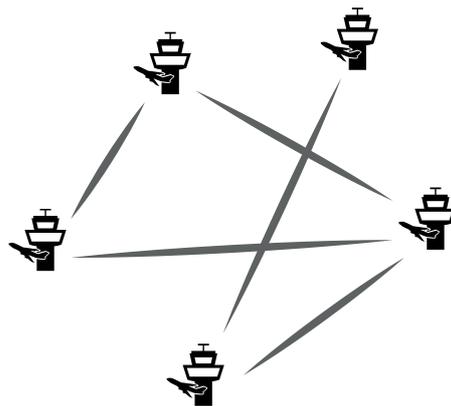

**Figure 1. Illustration of a simple undirected static airport network.** Towers and aircrafts represent airports, and lines represent the links between airports, i.e. flights between pairs of airports. The spatial position of airports in the network representation does not necessarily correspond to actual geographic locations.

A fundamental quantity of a network is the number of links, also called degree, of each node; it is given by $k_i = \sum_{j=1}^{N} a_{ij}$. Similarly, the strength of a node is given by

$s_i = \sum_{j=1}^{N} w_{ij}$ [8,9]. For directed networks, one also defines in-degree and out-degree to represent incoming and out-coming links, respectively. These quantities not only indicate the number of secondary nodes (i.e. the node's neighbors) or load (i.e. the node's weight) associated to a node but also gives the centrality of the node in the network. Centrality is a measure of the importance of the node and the degree, or the strength, measures how important the node is regarding its local connectivity. In other words, high degree means that a highly connected airport (i.e. a hub) is more important than one poorly connected. This measure is local and thus misses the fact that some nodes may connect different parts of the network even if they have relatively low degree. A typical example is an airport with several connections between two distinct continents (or regions of a country) even though this particular airport is not a major international hub. There are different ways of estimating this type of centrality [8,9]; the most common within air transport research is the betweenness centrality [8,9]. Betweenness $B_u$ measures the fraction of shortest paths between any two network nodes $i$ and $j$ passing through node $u$, divided by the number of possible shortest paths between all pairs $i$ and $j$. A path between nodes $i$ and $j$ is defined as the distance between the two nodes in terms of links traversed [8,9]. Consequently, high betweenness reflects the fact that a node is a bottleneck between two or more parts of the network. Other relevant centrality measures include the random walk and Page-rank [8,9].

The clustering of nodes can be also estimated by different measures. For example, the clustering coefficient of node $i$ measure the probability that two neighboring nodes say $u$ and $v$ (i.e. $a_{iu} = 1$ and $a_{iv} = 1$) are also linked (i.e. $a_{uv} = 1$) between themselves [8,9]. In other words, it is a measure of the number (or density) of triangles in the network. At the mesoscopic level, one may define the network community structure to identify groups of nodes more connected between themselves than with nodes at other groups [21]. A typical example is the community structure formed by a country in which several connections occur between airports within the country but few connections link a couple of international airports with their counterparts in another country. More details on these and other popular measures (e.g. assortativity or reachability) can be easily found in the literature, see for example Refs. [8,9].

**2.2. Temporal Networks**

A temporal network differs from a static network in the sense that now a different adjacency matrix is defined at each time $t$ [22], i.e. $\boldsymbol{A}(t)$ and consequently $a_{ij}(t)$ (or the equivalent for weighted versions of the network). The standard procedure is to collect all links within a time window $[t, t + \delta t)$ into a network snapshot (Figure 2). The parameter $\delta t$ can be varied from minutes to years, depending on the resolution available from data or on the research questions. For example, daily variations of flights may be irrelevant if one wants to map the long-term evolution of the air transport infrastructure in a given country. On the other hand, high-temporal resolution (i.e. small $\delta t$) becomes important if one wishes to understand the cascades of delays caused by unexpected weather conditions during a single day. The temporal framework becomes interesting when the scale of the processes taking place on the network is smaller (or at same order) than the scale of variations in the network structure. Under these conditions, it becomes relevant to measure characteristics of the network taking into account the temporal dynamics.

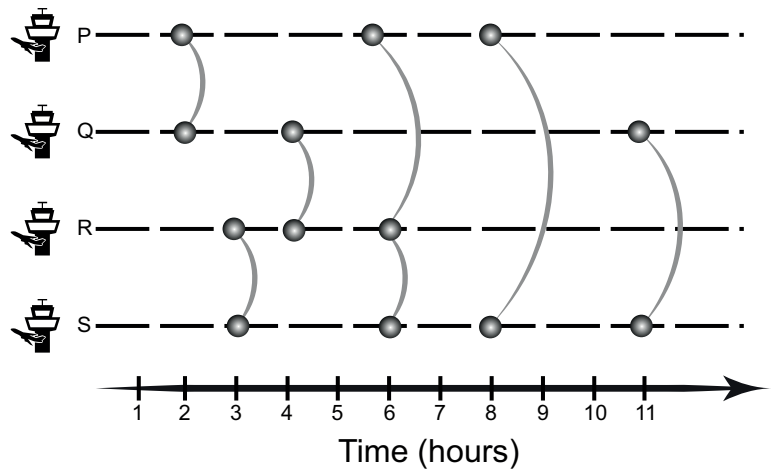

**Figure 2. Illustration of an airport temporal network.** Each dashed horizontal line represents a different airport (P, Q, R and S). Curved vertical lines represent active links between airports, i.e. a flight connecting two airports. The resolution here is $\delta t = 1$ hour, therefore, each snapshot aggregates links within one hour.

In the temporal framework, it is possible to either measure static network structures for each snapshot, i.e. for each $A(t)$, or to study the network as a continuous temporal sequence of paths. In the second approach, one considers only a single link per time step. For example, in order to move from airport P to R in Figure 2, one has to move from P to Q at time 2 and then, from Q to R at time 4. In this simple example, the topological distance is 2 but the temporal distance is 3 hours. Temporal paths typically reduce the number of potential paths and increase the distance between two nodes [23,24]. Figure 3 shows another example; although node P is central in the static framework (Figure 3a), it is not if the temporal paths are taken into account (Figure 3b) because some flights occur before others and the network becomes not strongly connected (i.e. not all nodes can be reached by any other node in the network). Note that in this simple example, the centrality of a node may completely change in a way not captured by using the standard static network formalism over different network snapshots.

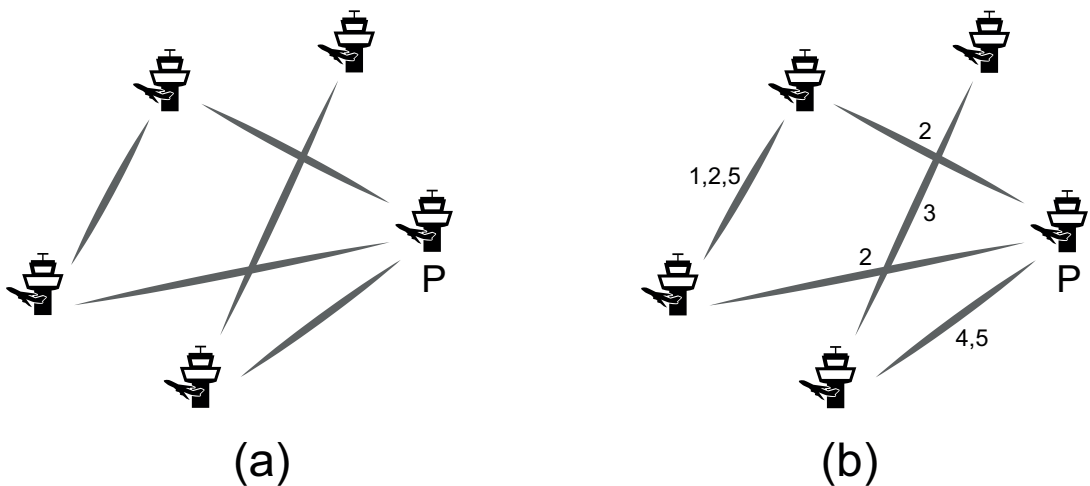

**Figure 3. Static vs. temporal centrality.** (a) Illustration of the static network representation of an airport network. Airport P is the most central. (b) Illustration of the temporal network representation of an airport network. Numbers represent the times in which the links are active or available. Airport P looses importance if considering the temporal paths.

One complication of modeling airports or flights as temporal networks is that a flight departs from airport P, say at time $t$, and arrives at another airport Q at time $t + \tau$. In principle, at times closer to $t$, this out-link is not "felt" by airport Q whereas at times closer to $t + \tau$, the in-link is not "felt" by airport P. In other words, because of the duration $\tau$ of the flight, the activation of the link is not synchronized in both airports, a phenomenon generally not observed in other systems modeled by temporal networks [22]. Although relevant at hourly or minute scales, this effect is irrelevant if modeling the network at lower temporal resolutions, as for example, daily or annually. Note however that this does not impede the propagation of delays for example, meaning that a departure 10 minutes late may imply on arrival 10 minutes late as well. More details on temporal networks, algorithms and measures can be found in the literature, see for example ref. [22].

### 2.3. Flight Data

Air transport data are widely available through different online sources. Nevertheless, many times a financial agreement has to be done in order to use them. Data at country level is typically easier to obtain than worldwide data however several times the first lacks structure and standardization for automatic downloading. Table 1 reports the sources of data used in the reviewed papers whenever it was clearly reported. In several studies, the data sources were not available or multiple sources were used for different years.

| Source | Reference |
|---|---|
| Civil Aviation Administration of China (CAAC) | 39, 29, 32 |
| Official Airline Guide – www.oag.com | 28, 37 |
| www.airdi.net | 25 |
| Bureau of Transportation Statistics – www.bts.org | 24, 42, 13, 34 |
| Air Traffic Management Bureau (ATMB) of China | 45 |
| Brazilian National Agency of Civil Aviation – www2.anac.gov.br/estatistica/anuarios.asp | 36 |
| United States Department of Transportation Federal Aviation Administration Form 5010 – www.gcr1.com/5010web | 35 |
| Traffic Flow Management System to Aircraft Situation Display to Industry Interface Control Document for the Traffic Flow Management Modernization Program – Federal Aviation Administration 2009 | 40 |

**Table 1.** Data sources as reported in reviewed papers.

### 2.4. Computational Tools

There are several computational tools to analyze complex network data. Some are more complete than others, and sometimes an implementation of a particular algorithm is available in the author's webpage. Those interested on a quick start to perform network analysis of air transport data but are unwilling to implement their own codes may look at Table 2 for a range of free stand-alone softwares or network libraries for known programing languages such as C/C++ (see e.g. gcc.gnu.org), R (www.r-project.org), Python (www.python.org) or Matlab® (www.mathworks.com).

| Software or library | Web-address |
|---|---|
| **Igraph** – Generate networks, analysis, etc. For C/C++, R, Python. | http://igraph.org/ |
| **Networkx** – Generate networks, analysis, etc. For Python. | http://networkx.github.io/ |
| **Pajek** – Analysis and visualization. Stand-alone software. | http://mrvar.fdv.uni-lj.si/pajek/ |
| **Cytoscape** – Analysis and visualization. Stand-alone software. | http://www.cytoscape.org |
| **Gephi** – Analysis and visualization. Stand-alone software. | https://gephi.org/ |
| **Fitting power-law distributions**. For C, R, Python, Matlab. | http://tuvalu.santafe.edu/~aaronc/powerlaws/ |
| **MapEquation** – Detect community structure and visualization. Stand-alone applet. | http://www.mapequation.org/apps.html |
| **Louvain method** – Detect community structure. For C++ and Matlab. | https://perso.uclouvain.be/vincent.blondel/research/louvain.html |

**Table 2.** Stand-alone software and libraries for network data analysis.

## 3. Airport Networks

In airport networks, nodes represent airports and links represent the fact that at least one flight exists between two airports. In some studies, airports serving the same city are collapsed into a single airport to represent the city. This section reviews fundamental and network-based properties of evolving airport networks.

### 3.1. Fundamental Properties

The long-term evolution of airport networks is a consequence of a number of factors, as for example, increase of gross domestic product [25], population growth, optimization of resources, profit, and governmental market regulations [26]. There is agreement that the deregulation of air transportation in the 1970s in the USA caused major impact on the establishment of flight routes [27]. Similar governmental actions affected the European and Chinese markets a decade later. One important consequence of deregulation is that airport networks mostly moved from point-to-point towards hub-spoke systems. As pointed out in ref. [28], hub-spoke routes increased 66% while spoke-spoke and hub-hub increased 55% and 26% respectively in Europe in the period 1990-1998. A consequence of this effect was a relative increase in the frequency of flights between hub-spoke airports followed by a decrease in hub-hub and spoke-spoke routes [28]. The network consequence is that some major airports became even more important to define the resilience of the airport system against failures.

In this period following deregulation of air transport in various countries, different markets showed a strong variation in the number of airports served by scheduled flights. In China, several studies identified that the number of airports substantially increased after the 1980s. For example, from 60 airports in 1984, China moved to 91 airports in 2006 [29] (Note that these numbers depend on the dataset used). An analysis, dating back to 1930, shows that the number of airports increased nearly linearly in China during the 20th century [30, 31]. The passenger traffic, on the other hand, increased 40 times during the same period [29], typically following the increase in the GDP [30, 32]. Similar growth was observed in the US airport network, in which airports increased from ~300 in 1990 to ~900 in 2010 (Note that these numbers also depend on the dataset used) and the number of transported passengers rose by nearly 50% during the same period [33-35]. In contrast to these trends, the Brazilian airport network has decreased in the period from 1995 to 2006, respectively from 211 to 142 airports [36]. The average degree also decreased in the same period in Brazil [36] whereas the growth in the number of connections seemed to stabilize in China after the mid-90s [32 29]. In Europe and worldwide, the number of connections per airport has also increased during the 1990s and 2000s [29, 37]. Various studies reported strong variation in the number of flights, passengers, cargo and mail in different routes, particularly from the perspective of stable and new cities/airports [36, 32, 34, 38].

It was generally observed a densification, in terms of the clustering coefficient, of the airport networks in China [39], whereas in both Brazil and the USA, a small decrease occurred after the 1990s [36, 31]. Mehta and colleagues reported a relatively small clustering coefficient for the US airport network if the network is considered at a daily resolution [40]. Nevertheless, all networks shown small-world characteristics, i.e. high clustering coefficient and small average shortest paths. The concentration of links also varied between low-cost and full-cost carriers, with a significant variation between winter and summer months and over the 2000s, as reported in the context of Portugal [41].

One important analysis from the complex network point of view is the degree distribution. The degree distribution is a histogram of the number of airports (or cities) with a given number of connections. All studied networks in the literature showed a right-skewed distribution of degree, meaning that a few airports (i.e. hubs) have several routes and flights whereas the majority of airports have only a few routes and flights. One attempt to characterize these distributions is to statistically fit a function to the empirical data. There is an ongoing debate on which is the most appropriate functional form to fit such distributions irrespective of the particularities of each context. From the qualitative perspective, the important feature is the broad distribution and not the best fit to data (See "fitting a power-law distributions" in Table 2). The analysis of the Brazilian [36], Chinese [32, 39] and the USA [31,13] networks further showed that the shape of the distribution is conserved over periods of approximately 10-20 years (Figure 4), meaning that even with strong rewiring at the micro-level, the macroscopic characteristics of the network remain roughly the same (i.e. has the same functional form possibly with different parameters). Similar results are observed for modeling the airport network at the country-level [25] (see more in section 5). On the other hand, the analysis of the worldwide air transportation network showed that the slope of the degree distribution decreased, meaning that relatively more airports are observed with very large degree [37].

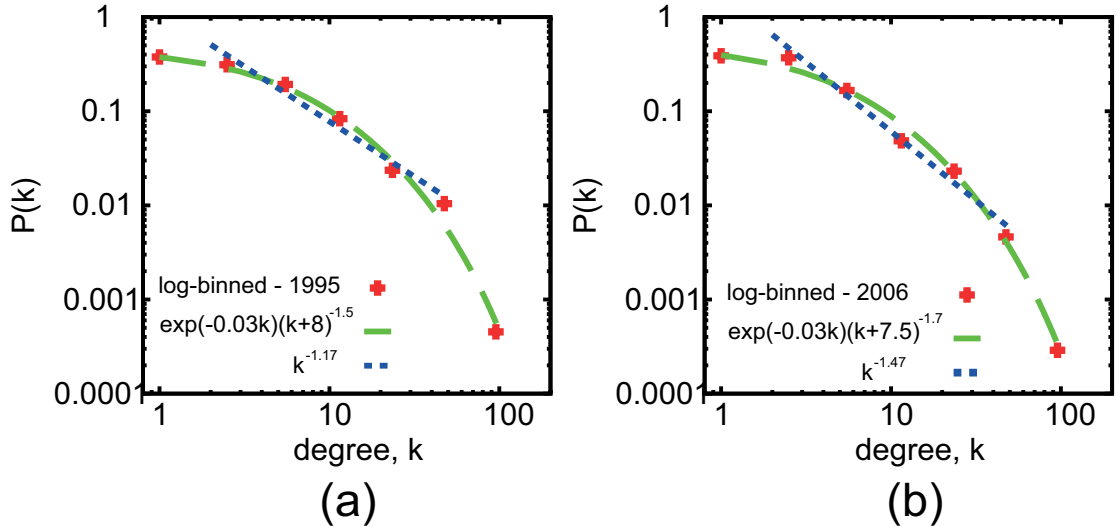

**Figure 4. Degree distribution of the Brazilian airport network**. Degree distribution in year (a) 1995 and (b) 2006. Both power-law (in blue short-dashed) and power-law with exponential cutoff (green long-dashed) functions were used to fit the logarithmic binned empirical data (red crosses). See ref. [36] for the data.

It is typically relevant to determine the importance of airports beyond the degree centrality. Such analysis helps to identify airports that are important in terms of their topological positions in the network. A widely used measure is the betweenness centrality (see section 2.1). The Brazilian airport network was the first study investigating the annual evolution of the betweenness for different airports. It was found that in a period of 10 years, within the top 10 airports in Brazil, betweenness increased almost 200% for one airport whereas it decreased 61% for another. All airports showed a strong variation in these values [36]. Generally speaking, the average betweenness centrality increased in China from 1930 to 1970 and then maintained relatively stable values in the following years [30]. The distribution of betweenness values for each airport also presented a broad distribution [31], a sign that a few airports are much more central (and critical) than the majority of them.

Bonnefoy and Hansman also studied the airport network of the USA by differentiating types of aircrafts. Unsurprisingly, routes involving turboprops and light piston mostly generate spatial networks where links have short spatial coverage whereas wide and narrow body jets created the typically observed long-range connections between far airports [35] as those described above.

### 3.2. Community Structure

The pattern of connections between airports in a given period of time may imply that airports are clustered in groups, i.e. network communities. These groups are not necessarily stable but evolve with the network. Gegov and colleagues performed a network community analysis on the USA airport networks by looking at a sequence of networks aggregated every 2 months [42] in the USA context. They have observed that community structure generally overlap with spatial distribution but some airports may not be in the same area (e.g. west coast and northwest form a single community, that is more explicit and stable in 2000 than in 1990, but not in 2010). A variation of membership on each community or group also occurs within a year. Interestingly, they have found a strong correlation between air travel within the same (and between)

community(ies) and migration patterns. The evolution of the community structure was also briefly investigated in ref. [43].

**3.3. Other Analysis**

Sophisticated methods, beyond betweenness, to analyze the connectivity and node centrality are available in the literature. For example, Allroggen and collaborators [44] devised two indexes, named global connectivity and global hub centrality, to measure the quality of scheduled air services taking into account frequency, detours, layover time and destination quality derived from passenger behavior. The goal was to develop a centrality measure taking into account quality rather than shortest paths between airports. They have analyzed the worldwide network from 1990 to 2012 and observed a growth of connectivity but heterogeneous trends in terms of type of connectivity. The USA and European airports accounted for 63% and 23% of global nonstop connectivity in 1990. This connectivity share of airports declined from 92.0% to 75.4% in the 2000s. In 1990, 87.4% of global one-stop connectivity was facilitated through hub operations in the USA. Between 1990 and 2000, 93.6% of the global hub centrality growth occurred at airports in North America and Europe. In particular, strong centrality growth of European airports caused the global hub centrality share of European (USA) airports to increase (decrease) from 10.0% (87.4%) in 1990 to 20.1% (75.3%) in 2000.

**4. Air transport country network**

In order to understand the role of countries in the global air transportation network, it may be useful to group, into a single node, all airports of a given country such that connections only occur between different countries. This course-grained approach helps to identify mobility patterns between countries. Wandelt and Sun [25] showed that the average degree increases from nearly 35 (in 2002) to nearly 40 (in 2014). As expected, the average degree suffers a cyclic pattern in which summer months contain more routes than winter months. They have also applied a number of network measures to identify critical nodes, that is, the USA, France, Great Britain, Australia and South Africa (with United Arab Emirates increasing its overall importance). Critical links on the other hand involved Great Britain to USA, Japan to USA, and France to Great Britain.

**5. Air route network**

Similarly to highways or railways, air transport is also limited by airways, that is, predefined regions of the sky in which civil aircrafts are allowed to fly. Airways not necessarily follow the shortest path between two airports but they simplify and facilitate control by radar operators. Kai-Quan and colleagues studied the air routes (or airways) of China [45]. As expected, airways are strongly constrained by spatial patterns and as a consequence the network has 1,013 nodes (in contrast to the airport network with 147 according to the same study) and 1,586 links, with a relatively low average degree of 3.13. The spatial constrains also imply a large diameter of 39, very small clustering coefficient (0.08) and low betweenness. The degree (and strength) distribution is closer to an exponential distribution, a direct consequence of the spatial nature of the network, meaning that no nodes are favored and a characteristic number of links is observed. The shape of the distribution remains the same from 2002 to

2010 but the exponent decreases indicating an increase in the traffic flow. In fact, traffic flow increased exponentially in this period.

## 6. Temporal Networks

A particular characteristic of the previous studies is that they estimate network statistics using snapshots of the network at a given time or interval of time, typically of at least one day. This approach means that one performs a static network analysis at different times and studies how these measures evolve in time. Nonetheless, when processes taking place in the network occur at scales comparable to the variations in the network structure, it becomes relevant to measure characteristics of the network taking into account all temporal paths simultaneously (see Section 2.2). In this section we review state-or-art results on airport networks using temporal paths and time-series analysis of airport activity.

### 6.1. Temporal Paths

Pan and Saramäki introduced the concept of path lengths and centrality in temporal networks (see also Ref. [23]), and applied their methods to the USA airport network during 10 subsequent days in 2008 [24]. They have observed that nodes relatively close in the static network may be connected via slow (farther) paths in the temporal version of the network. Moreover, correlations and heterogeneities in the event sequences, due to optimized scheduling of flights, imply that temporal path lengths decrease in this network.

### 6.2. Delays

Depending on the research questions, the study of temporal airport networks may also benefit of tools from time-series analysis. By neglecting the network structure, one may focus on the temporal aspects of the evolution of some variable, as for example, the number of flights or accumulated delay in a given airport. Such perspective may benefit of a vast literature on time-series analysis [46], including methods for forecast behavior, synchronization of activity, short or long-term correlations, and Fourier analysis. One such study, by Belkoura and Zanin [47], involves the analysis of the delay propagation on the 100 busiest airports in Europe during a few months in 2011. The average delay is calculated using 1-hour time intervals for each airport. By using the Granger causality (to analyse systemic delays) and a metric for extreme events, the authors concluded that systemic and extreme delays propagate in different ways within such network of airports [47]. Both the threshold (for delay propagation) and the additional delay, needed to trigger a phase change, increase with the airport traffic. They have also identified that large airports cause less delay propagation than smaller airports, possibly because of their available resources to manage critical situations [47, 48].

In the context of the USA, Fleurquin and collaborators observed that in 2010, delays showed similar statistics for different days of the week and different seasons. Major differences however were observed for different airports. In all cases, broad distributions of delays were observed. For long delays (more than 12 hours), they observed a relative concentration of delays early in the morning and late in the afternoon, in contrast to short delays (less than 12 hours) that showed a nearly flat delay distribution [49] during business hours.

## 7. Conclusions and Perspectives

Air transport systems play a crucial role in human mobility, transportation of goods, and spread of infectious diseases. Understanding the mechanisms driving air transport dynamics is a fundamental step to better control and optimize such systems. Air transport systems are also highly dynamic, the long-term evolution reflects population and economic growth, and the short-term dynamics reflect passenger needs and optimization of resources. Here, we briefly review studies dealing with the evolution of air transport systems at different temporal and spatial scales. The goal is to map the state-of-art to identify the advances in the field and areas in need of further research.

Generally speaking, we have observed that significant research has been done at the annual, monthly and daily scales taking the perspective of evolving networks as subsequent snapshots of static networks, that is, each snapshot is analyzed using statistics for static networks and the evolution of the variable of interest is considered. Nevertheless, little has been done to understand how evolving networks affect the spread of infections or the resilience of the air transport network against failures and targeted attacks. Furthermore, only a few studies approach evolving networks from the perspective of temporal paths. The study of temporal paths has gained momentum in recent years within the broader field of network science, particularly on the topic of diffusion processes on dynamic networks [50-52]. The biggest advantage of temporal paths is that they capture the relation between subsequent snapshots of the network in contrast to previous approaches in which this inter-snapshot relation was discarded or neglected. Temporal paths are particularly relevant to study highly dynamic structures, as for example, the hourly dynamics within a single day. Relevant problems not yet fully studied include the effect of evolving structures on delays and consequently the resilience of the network during a given day. Not less important is how to improve connectivity aiming to decrease total travel time while optimizing resources (i.e. aircrafts, airport slots and crew) and maximizing profits. Geographic constrains may also play a role; understanding its importance together with the temporal aspects of flights remain an open problem.

As mentioned above, research on temporal paths (broadly known as temporal networks) is still in its infancy [22]. Nevertheless, several studies already provide useful tools that could be applied to the understanding of air transport systems. For example, centrality measures (see section 4.4 of ref. [22] for references), the characterization of temporal paths (see section 4.3 of ref. [22]), community structure [53], or flow motifs (see section 4.10 of ref. [22]). The caveat however is that such methods have to be adapted to the asynchronous nature of flights at different airports, as discussed in section 2.2 of this review. One possibly fast track is to use methods based on diffusion processes, as for example random walks or epidemics, because in these cases it is relatively easier to deal with asynchronous connections, that is, a walker (or an "infection") may leave a node at a given time $t$ and arrive at another node at time $t + \tau$. Irrespective of the methodology, further collaborations between air transport experts and network scientists are necessary to make sense of network measures in the context of air transport.

**Acknowledgements**

This work was supported by the Fonds de la Recherche Scientifique - FNRS.

**Luis E C Rocha** is Assistant Professor at Karolinska Institutet, Sweden, and FNRS Research Fellow at the Université de Namur, Belgium. He has a background in computational physics and complex systems. His research is about dynamics on and of networks, i.e. the study of dynamic processes such as epidemics and random walks on networks that vary in time. His expertise is on computational modeling and data analysis of theoretical and real-life networks particularly on networks of relevance to public health. He is also interested on human mobility and transportation, particularly on how they regulate diffusion processes on networks.